\documentclass[prl,twocolumn,showpacs,superscriptaddress,preprintnumbers]{revtex4}
\usepackage{amssymb}
\usepackage{graphicx}
\usepackage{dcolumn}
\usepackage{bm}
\usepackage[bookmarks = true, pdfpagemode = None, pdfstartview = FitH,colorlinks = true,
citecolor = black, linkcolor = black, urlcolor = black]{hyperref}
\usepackage{url}

\newcommand{\bra}[1]{\left\langle{#1}\right\vert}
\newcommand{\ket}[1]{\left\vert{#1}\right\rangle}

\begin{document}

\title{Synchronization of Multiple Coupled rf-SQUID Flux Qubits}
\author{R.~Harris}
\email{rharris@dwavesys.com}
\affiliation{D-Wave Systems Inc., 100-4401 Still Creek Dr., Burnaby, BC V5C 6G9, Canada}
\homepage{www.dwavesys.com}
\author{F.~Brito}
\author{A.J.~Berkley}
\author{J.~Johansson}
\author{M.W.~Johnson}
\author{T.~Lanting}
\author{P.~Bunyk}
\author{E.~Ladizinsky}
\affiliation{D-Wave Systems Inc., 100-4401 Still Creek Dr., Burnaby, BC V5C 6G9, Canada}
\author{B.~Bumble}
\author{A.~Fung}
\author{A.~Kaul}
\author{A.~Kleinsasser}
\affiliation{Jet Propulsion Laboratory, California Institute of Technology, Pasadena CA, USA}
\author{S.~Han}
\affiliation{Department of Physics and Astronomy, University of Kansas, Lawrence KS, USA}

\date{\today }

\begin{abstract}
A practical strategy for synchronizing the properties of compound Josephson junction rf-SQUID qubits on a multiqubit chip has been demonstrated.  The impacts of small ($\sim1\%$) fabrication variations in qubit inductance and critical current can be minimized by the application of a custom tuned flux offset to the CJJ structure of each qubit.  This strategy allows for simultaneous synchronization of the qubit persistent current and tunnel splitting over a range of external bias parameters that is relevant for the implementation of an adiabatic quantum processor.

\end{abstract}

\pacs{85.25.Dq, 03.67.Lx}
\maketitle

Despite daunting obstacles, there is considerable interest in the development of solid state quantum information processors.  This interest is fueled by the hope that breakthroughs in device fabrication will eventually facilitate the realization of large scale quantum processors whose performance could surpass that of classical computers.  Implementations based upon superconducting qubits have received particular attention \cite{SCQubits}.  Considerable effort has been made in studying noise in such circuits \cite{Noise,Drift}.  An equally pressing matter is fabrication variability as qubits are acutely sensitive to variations in device parameters \cite{Sensitivity,CJJ}.  Current state of the art superconducting fabrication technology is limited, at best, to $ \sim 1\%$ spreads in parameters such as Josephson junction critical currents and qubit inductances. The extent to which this variability affects the performance of a superconducting quantum computer is an important open problem.
Therefore, it is relevant to demonstrate superconducting qubit designs and methods of operation that are insensitive to small variations in device parameters.

We wish to focus on a quantum Ising spin glass simulator \cite{Farhi,AQC} constructed from superconducting flux qubits \cite{architecture}.  Such a device could be useful for solving optimization problems \cite{Boros}.  Each qubit $i$ serves as a spin-$1/2$ subjected to transverse and longitudinal biases $\Delta_i$ and $\epsilon_i\equiv\mu_iB_i$, respectively.  Here, $\mu_i$ represents the effective magnetic moment and $B_i$ an externally controlled magnetic field.  Pairwise couplings are realized by $J_{i,j}=M_{i,j}\mu_i\mu_j$, where $M_{i,j}$ is an externally controlled parameter.  The system Hamiltonian at any time during operation has the form
\begin{equation}
\label{eqn:HSpin}
{\cal H}=-\sum^N_{i=1}\frac{1}{2}\left[\epsilon_i\sigma_z^{(i)}+\Delta_i\sigma_x^{(i)}\right]+\sum_{i<j}J_{i,j}\sigma_z^{(i)}\sigma_z^{(j)}   \;\; .
\end{equation}

\noindent A particular adiabatic quantum algorithm, such as that described in Ref.~\cite{Farhi}, may require all $\mu_i$ and $\Delta_i$ to be nominally equivalent between qubits.  While this choice of algorithm is by no means unique, it does represent the simplest implementation of an optimization procedure that utilizes quantum adiabatic evolution.  The objective of the work presented herein was to develop a practical strategy for minimizing the differences in qubit parameters between superconducting flux qubits due to fabrication variations via in-situ tunable device biases.  

\begin{figure}[tbp]
\includegraphics[width=2.25in]{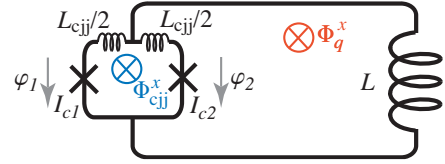}
\caption{(Color online) Schematic diagram of a CJJ rf-SQUID.}
\label{fig:CJJRFS}
\end{figure}

One useful implementation of a superconducting flux qubit is the compound Josephson junction (CJJ) rf-SQUID \cite{CJJ}, as depicted in Fig.~\ref{fig:CJJRFS}.  Here,  a main loop of superconducting wire of inductance $L_q$ is interrupted by a smaller loop of inductance $L_{\text{cjj}}$ with two Josephson junctions of critical current $I_{c1}$ and $I_{c2}$.  The CJJ and main loop are subjected to external fluxes $\Phi_{\text{cjj}}^x=\Phi_0\varphi_{\text{cjj}}^x/2\pi$ and $\Phi_q^x=\Phi_0\varphi_q^x/2\pi$, respectively ($\Phi_0\equiv h/2e$).  The Hamiltonian for this system can be written as
\begin{equation}
\label{eqn:2JHphase}
{\cal H}=\sum_{i=1}^2\left[\frac{Q_i^2}{2C_i}-E_{Ji}\cos(\varphi_i)\right]+\sum_{n}U_n\frac{\left(\varphi_n-\varphi_n^x\right)^2}{2}
\end{equation}

\noindent where $C_{i}$ and $E_{Ji}=I_{ci}\Phi_0/2\pi$ represent the capacitance and 
Josephson energy of junction $i$, respectively, and $[\Phi_0\varphi_i/2\pi,Q_j]=i\hbar\delta_{ij}$.  The inductive terms originate from the two closed loops with $n\in\left\{q,{\text{cjj}}\right\}$, $L_q\equiv L+L_{\text{cjj}}/4$ and $U_n\equiv(\Phi_0/2\pi)^2/L_n$.  The qubit and CJJ loop phases are defined as $\varphi_q\equiv\left(\varphi_1+\varphi_2\right)/2$ and $\varphi_{\text{cjj}}\equiv\varphi_1-\varphi_2$, respectively.  This 2-dimensional system can be reduced to an effective 1-dimensional Hamiltonian if $L_q\gg L_{\text{cjj}}$ because the plasma energy of the CJJ loop will be much higher than that of the main rf-SQUID loop.  Setting $\varphi_{\text{cjj}}=\varphi_{\text{cjj}}^x$ and combining the Josephson terms,
\begin{equation}
\label{eqn:2JHeff}
{\cal H}\approx \frac{Q_q^2}{2C_p}+V(\varphi_q)
\end{equation}
\vspace{-12pt}
\begin{displaymath}
V(\varphi_q)=U_q\Big\{\frac{\left(\varphi_q-\varphi_q^x\right)^2}{2}-\beta_{\text{eff}}\cos\left(\varphi_q-\varphi_q^0\right)\Big\}
\end{displaymath}
\vspace{-12pt}
\begin{displaymath}
\beta_{\text{eff}}=\beta_+\cos(\varphi_{\text{cjj}}^x/2)\sqrt{1+\left[\frac{\beta_-}{\beta_+}\tan(\varphi_{\text{cjj}}^x/2)\right]^2}
\end{displaymath}
\vspace{-12pt}
\begin{displaymath}
\varphi_q^0\equiv 2\pi\frac{\Phi_q^0}{\Phi_0} =-\arctan\left(\frac{\beta_-}{\beta_+}\tan(\varphi_{\text{cjj}}^x/2)\right)
\end{displaymath}

\noindent where $C_p\equiv C_1+C_2$, $[\Phi_0\varphi_q/2\pi,Q_q]=i\hbar$ and $\beta_{\pm}\equiv 2\pi L_q\left(I_{c1}\pm I_{c2}\right)/\Phi_0$.  Focussing upon the two lowest lying states in the regime $\beta_{\text{eff}}\lesssim -1$ ($\pi$ phase shifted flux qubit), one can recast Eq.~\ref{eqn:2JHeff} as a qubit Hamiltonian ${\cal H}_{q}=-{\frac{1}{2}}\left[\epsilon\sigma_z+\Delta\sigma_x\right]$,
where $\epsilon=2\left\vert I_q^p\right\vert\left(\Phi_q^
x-\Phi_q^0\right)$.  Denoting the ground and first excited state of Eq.~\ref{eqn:2JHeff} at $\Phi_q^x=\Phi_q^0$ by $\ket{+}$ and $\ket{-}$, respectively, the spin
states can be expressed as $\ket{\uparrow}=\left(\ket{+}+\ket{-}\right)/\sqrt{2}$
and $\ket{\downarrow}=\left(\ket{+}-\ket{-}\right)/\sqrt{2}$.
The persistent current is then defined by $|I_q^p|\equiv\left|\bra{\uparrow}\left(\Phi_q-\Phi_q^0\right)/L_q\ket{\uparrow}\right|$.  The tunneling energy is given by $\Delta=\bra{-}{\cal H}\ket{-}-\bra{+}{\cal H}\ket{+}$.  

The CJJ rf-SQUID was first proposed as a means of providing in-situ tuning of the tunnel barrier $U_q\beta_{\text{eff}}$ via $\Phi_{\text{cjj}}^x$ \cite{CJJ}.  By allowing for small relative offsets in the CJJ bias $\delta\Phi_{\text{cjj}}$, it is possible to simultaneously minimize differences in $\left|I_q^p\right|$ and $\Delta$ between qubits with slightly different $L_q$ and $I_c\equiv I_{c1}+I_{c2}$, thus {\it synchronizing} their properties.  Consider $\left|I_q^p\right|$ and $\Delta$ in the regime $U_q\beta_{\text{eff}}\gg\hbar\omega_p\equiv \hbar/\sqrt{L_qC_p}$.  In this scenario, $\left|I_q^p\right|$ is primarily determined by the position of the minima of $V(\varphi_q)$ with only an extremely weak dependence upon $C_p$.  In order to maintain constant $\left|I_q^p\right|$ in the presence of small variations in $\alpha\in \left\{L_q,I_c\right\}$, the condition is $\beta_{\text{eff}}(\alpha,\Phi_{\text{cjj}}^x)=\beta_{\text{eff}}(\alpha+\delta\alpha,\Phi_{\text{cjj}}^x+\delta\Phi_{\text{cjj}})$.  To first order in $\delta\alpha/\alpha$, $\delta\Phi_{\text{cjj}}\approx(\Phi_0/\pi)\left[\cot\left(\pi\Phi_{\text{cjj}}^x/\Phi_0\right)\right]\delta\alpha/\alpha$.  For $\left|\delta\alpha/\alpha\right|= 0.05$, one obtains $\left|\delta\Phi_{\text{cjj}}\right|\sim15\,$m$\Phi_0$.  Furthermore, one can use the WKB approximation \cite{Landaubook} to write
\begin{equation}
\label{eqn:WKBDelta}
\Delta\approx\frac{\hbar\omega_p}{\pi}e^{-\frac{\Phi_0}{2\pi\hbar}\sqrt{2 C_p}\int_{-a}^a\! d\varphi_q\sqrt{V(\varphi_q)-\hbar\omega_p}}
\end{equation}

\noindent where $\pm a$ represent the classical turning points straddling the local maximum in $V(\varphi_q)$.  The resultant form for $\Delta$ reveals that $\Delta(\alpha,\Phi_{\text{cjj}}^x)=\Delta(\alpha+\delta\alpha,\Phi_{\text{cjj}}^x+\delta\Phi_{\text{cjj}})$, where $\delta\Phi_{\text{cjj}}\approx\gamma(\Phi_0/\pi)\left[\cot\left(\pi\Phi_{\text{cjj}}^x/\Phi_0\right)\right]\delta\alpha/\alpha$, with $\gamma\sim 1$ for $\alpha=L_q,I_c$ and $\gamma\ll 1$ for $\alpha=C_p$.  Interestingly, $\Delta$ shows a relatively weak dependence upon $C_p$ as compared to $L_q$ and $I_c$.  Thus perturbations of $L_q$ and $I_c$ ($\lesssim 5\%$) result in approximately the same shift of the CJJ bias-dependence of both $\left|I_q^p\right|$ and $\Delta$.  In contrast, perturbations of $C_p$ ($\lesssim 5\%$) have negligible impact upon $\left|I_q^p\right|$ but do influence the CJJ bias dependence of $\Delta$.

The above observations indicate that one can compensate for small variations of $L_q$ and $I_c$ between CJJ rf-SQUID qubits by the application of custom tuned CJJ bias offsets.  For typical device parameters, $\Delta/h$ varies from $\sim 1\,$MHz to $\sim10\,$GHz for $1\lesssim\left|\beta_{\text{eff}}\right|\lesssim 1.3$: If the qubit has been designed with $\beta_+\gtrsim 1.5$, then the range of $\Phi_{\text{cjj}}^x$ that is relevant for operation will be $\lesssim50\,$m$\Phi_0$ wide.  Thus, one can choose a unique reference CJJ bias $\Phi_{\text{cjj}}^0$ in the center of the operating regime for each qubit such that $\left|I_q^p\right|(\Phi_{\text{cjj}}^x-\Phi_{\text{cjj}}^0)$ and $\Delta(\Phi_{\text{cjj}}^x-\Phi_{\text{cjj}}^0)$ are synchronized.

\begin{figure}[tbp]
\includegraphics[width=3.0in]{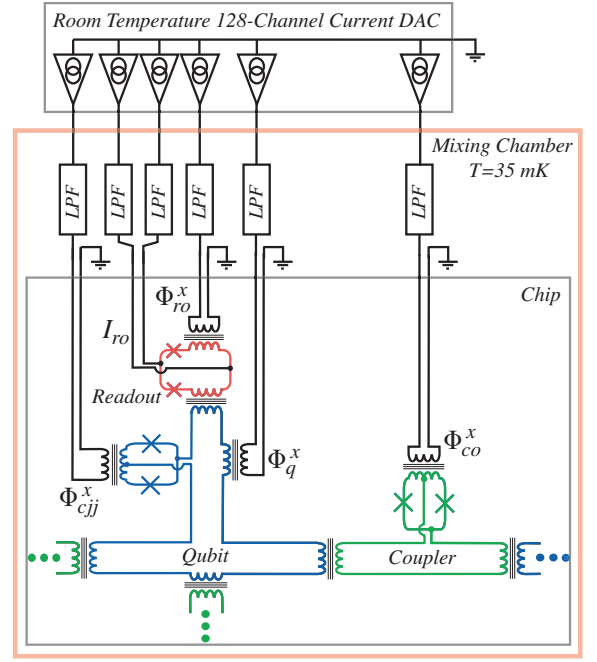}\\
\caption{(Color online) Schematic of a portion of a multiqubit chip, bias line configuration and room temperature electronics.  Ellipses indicate devices (alternating between qubit and coupler) that extend beyond the scope of the diagram.}
\label{fig:Schematic}
\end{figure}

In order to experimentally assess the CJJ synchronization strategy we focussed on a subset of CJJ rf-SQUID flux qubits embedded in a larger lattice of such devices (see Fig.~\ref{fig:Schematic}).  Each qubit is connected to three others via in-situ tunable monostable CJJ rf-SQUID couplers, which we treat as classical mutual inductances \cite{CJJCoupler}.  We isolated a linear chain of six qubits by setting the intervening couplers to maximum antiferromagnetic coupling and the remaining unused couplers to provide zero coupling.  Each qubit's state was probed via a dedicated dc-SQUID magnetometer.  The chip was fabricated from an oxidized Si wafer with Nb/Al/Al$_2$O$_3$/Nb trilayer junctions and three Nb wiring layers separated by sputtered SiO$_{2}$.  It was mounted to the mixing chamber of a dilution refrigerator regulated at $T=35\,$mK inside a PbSn superconducting magnetic shield with a residual field in the vicinity of the chip $\lesssim 9\,$nT.  External current biases were provided by a custom-built programmable room temperature 128-channel current DAC.  Low pass filters (LPFs) with $f_c\approx 5\,$MHz were constructed from a combination of lumped element and copper powder filters secured to the mixing chamber.  All mutual inductances and residual flux offsets were calibrated in-situ.

\begin{figure}[tbp]
\includegraphics[width=3.0in]{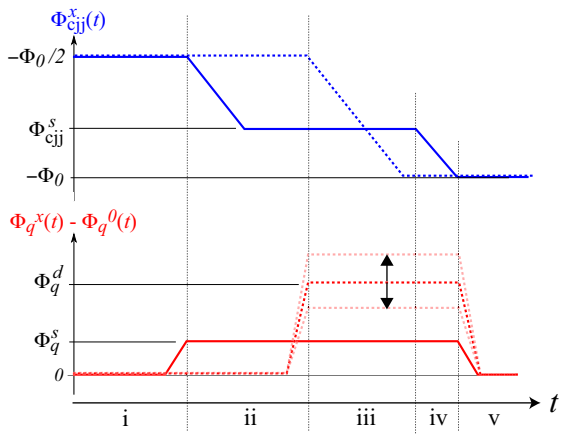}
\caption{(Color online) CJJ and flux bias waveforms versus time $t$ for source (solid) and detector (dashed) qubits.}
\label{fig:Waveforms}
\end{figure}

To measure $\left|I_q^p\right|$ we first employed dc-SQUIDs as magnetometers.  To begin, we initialized a given qubit, hereafter referred to as the source qubit, in the state $\ket{\uparrow}$, raised $U_q\beta_{\text{eff}}$ to maximum height ($\Phi_{\text{cjj}}^x=-\Phi_0$) and measured the change in flux sensed by its dc-SQUID.  The process was repeated for the qubit initialized in the state $\ket{\downarrow}$ and the difference between the two measurements recorded.  Knowing the readout to qubit mutual inductance $M_{\text{ro-q}}=6.46\pm0.17\,$pH, obtained from an independent measurement, we determined $\left|I_q^p\right|_{\text{max}}\equiv\left|I_q^p\right|(\Phi_{\text{cjj}}^x=-\Phi_0)$ for each qubit.  This measurement technique provided reliable results only if $M_{\text{ro-q}}\left|I_q^p\right|>\delta\Phi_{\text{ro}}$, where $\delta\Phi_{\text{ro}}\sim 2\,$m$\Phi_0$ represents a flux resolution limit imposed by the width of the dc-SQUID switching current distribution.  In order to clearly resolve $\left|I_q^p\right|$ with $U_q\beta_{\text{eff}}$ suppressed, we utilized a second qubit, hereafter referred to the detector qubit, that was coupled to the source qubit via a coupler with effective mutual inductance $M_{\text{eff}}$.  Referring to Fig.~\ref{fig:Waveforms}, the sequence began with both qubit $V(\varphi_q)$ monostable ($\Phi_{\text{cjj}}^x=-\Phi_0/2$) and biased to their degeneracy points ($\Phi_q^x-\Phi_q^0=0$) (i).  Next, the source qubit was partially annealed to an intermediate CJJ bias $-\Phi_0<\Phi_{\text{cjj}}^s<-\Phi_0/2$ in the presence of a small bias $\Phi_q^s=\pm2.1\,$m$\Phi_0$ in order to initialize its state (ii).  Thereafter, the detector qubit was fully annealed ($\Phi_{\text{cjj}}^x$ ramped to $-\Phi_0$) in the presence of a variable bias $\Phi_q^d$ (iii).  Finally, the source qubit was fully annealed (iv), both qubit flux biases are returned to their degeneracy point (v) and the state of the detector qubit is read (not shown).  This annealing cycle was embedded inside a software feedback loop which adjusted $\Phi_q^d$ until the particular bias for which the detector qubit could be found in $\ket{\uparrow}$ with probability $P_{\uparrow}=1/2$ was found to within a specified precision.  Performing the measurement for both signs of $\Phi_q^s$ and taking the difference between the two resultant values of $\Phi_q^d$ yielded $2M_{\text{eff}}\left|I_q^p\right|$.  Given $\left|I_q^p\right|_{\text{max}}$ we then inferred $M_{\text{eff}}=1.35\pm0.04\,$pH for the 5 intervening couplers in the chain of 6 qubits.  It was then possible to scale maps of $2M_{\text{eff}}\left|I_q^p\right|$ versus $\Phi_{\text{cjj}}^x$ to extract $\left|I_q^p\right|(\Phi_{\text{cjj}}^x)$.

To measure $\Delta$ of each qubit we used two methods:  In the incoherent regime one can utilize macroscopic resonant tunneling (MRT) to trace out decay rate curves and extract $\Delta$ from fitting parameters \cite{MRT}.  The range of $\Delta$ that could be probed by this method had a practical upper bound on account of the relatively low bandwidth of our bias lines.  In the coherent regime, we employed a 2-qubit procedure involving the waveform pattern shown in Fig.~\ref{fig:Waveforms} in which $\Phi_q^s$ was scanned through the domain $\left[-3,+3\right]\,$m$\Phi_0$ and $\Phi_q^d$ was again adjusted via a software feedback procedure to determine the shift in detector qubit degeneracy point at each $\Phi_q^s$.  Knowing $\left|I_q^p\right|(\Phi_{\text{cjj}}^x)$  allowed one to convert $\Phi_q^s$ and $\Phi_q^d$ into $\epsilon_1$ and $\epsilon_2$, respectively.  For two coupled qubits in the limit $\Delta_2\rightarrow 0$ the eigenenergies of Eq.~\ref{eqn:HSpin} are given by $E_{1\pm}=\frac{1}{2}\left[\pm F(-1)-\epsilon_2\right]$ and $E_{2\pm}=\frac{1}{2}\left[\pm F(+1)+\epsilon_2\right]$, where $F(x)\equiv\sqrt{\left(\epsilon_1+x 2J_{1,2}\right)^2+\Delta_1^2}$.  Using Boltzmann statistics, one can calculate the particular bias $\epsilon_2=\epsilon_2^*$ for which the detector qubit will be found with $P_{\uparrow}=1/2$:
\begin{equation}
\label{eqn:HalfCondGeneral}
\epsilon_2^*= \frac{F(+1)-F(-1)}{2}+k_BT\ln\left(\frac{1+e^{-F(+1)/k_BT}}{1+e^{-F(-1)/k_BT}}\right)
\end{equation}

\noindent  Note that in the limit $\Delta\gg T,J_{1,2}$ Eq.~\ref{eqn:HalfCondGeneral} reduces to $\epsilon_2^*\approx J_{1,2}\epsilon_1/\sqrt{\epsilon_1^2+\Delta_1^2}=J_{1,2}\bra{g}\left|I_q^p\right|\sigma_z\ket{g}$, with $\ket{g}$ representing the groundstate of the source qubit.  Given independent calibrations of $\left|I_q^p\right|$ for both qubits, $M_{\text{eff}}$ and $T$, one can fit traces of $\epsilon_2^*$ versus $\epsilon_1$ with Eq.~\ref{eqn:HalfCondGeneral} to extract $\Delta_1$.  This procedure is similar in spirit to that of Ref.~\cite{IPHTDelta}.  In practice, the 2-qubit method was found to be reliable only if $\Delta>T$ and $M_{\text{eff}}\left|I_q^p\right|\gg \delta\Phi_{n}$, where $\delta\Phi_{n}$ represents an rms low frequency flux noise experienced by the detector qubit.  These constraints imposed lower and upper bounds, respectively, upon the range of $\Delta_1$ that could be probed via this latter method.

\begin{figure}[tbp]
\includegraphics[width=3.3in]{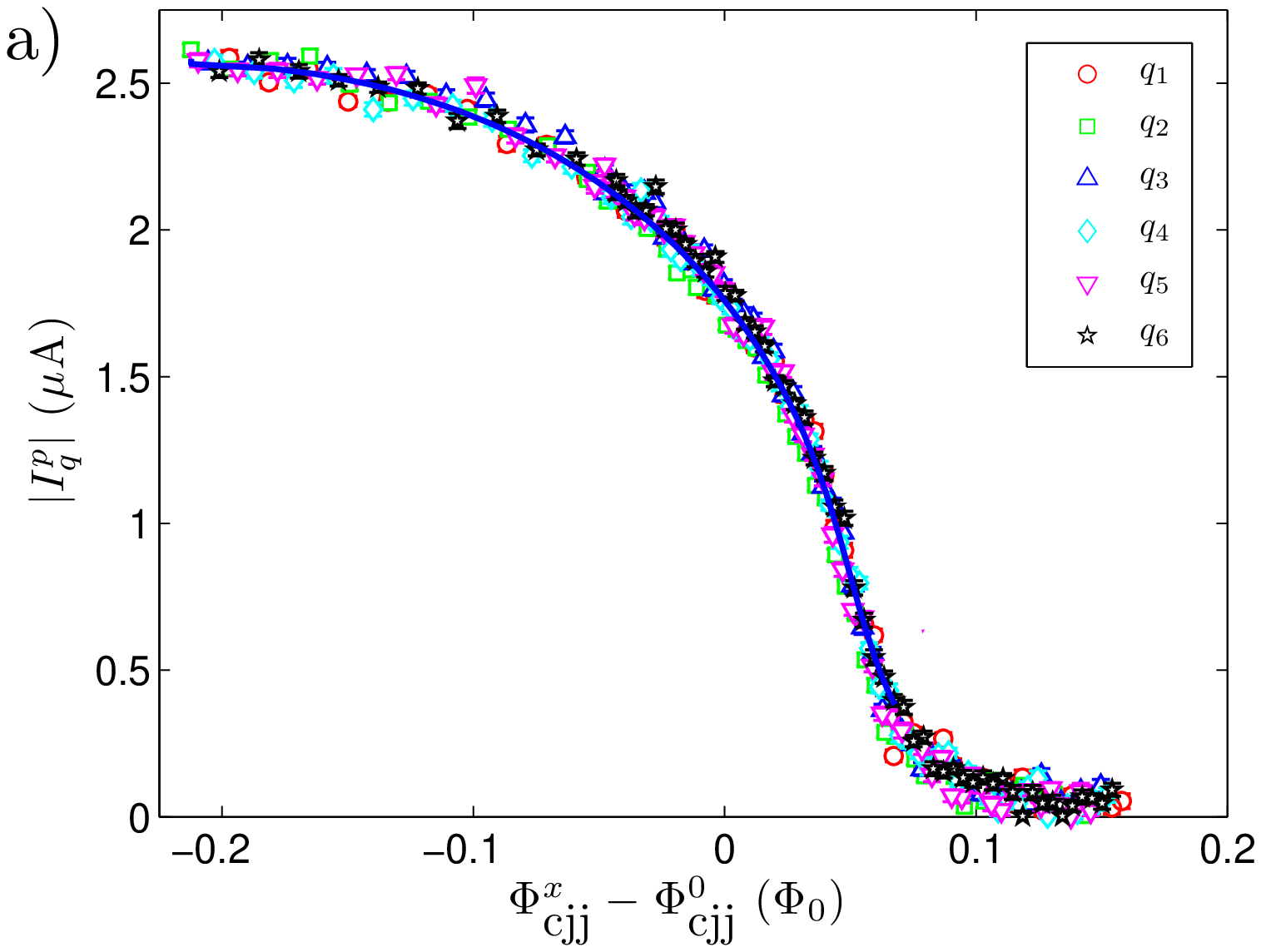}\\
\includegraphics[width=3.3in]{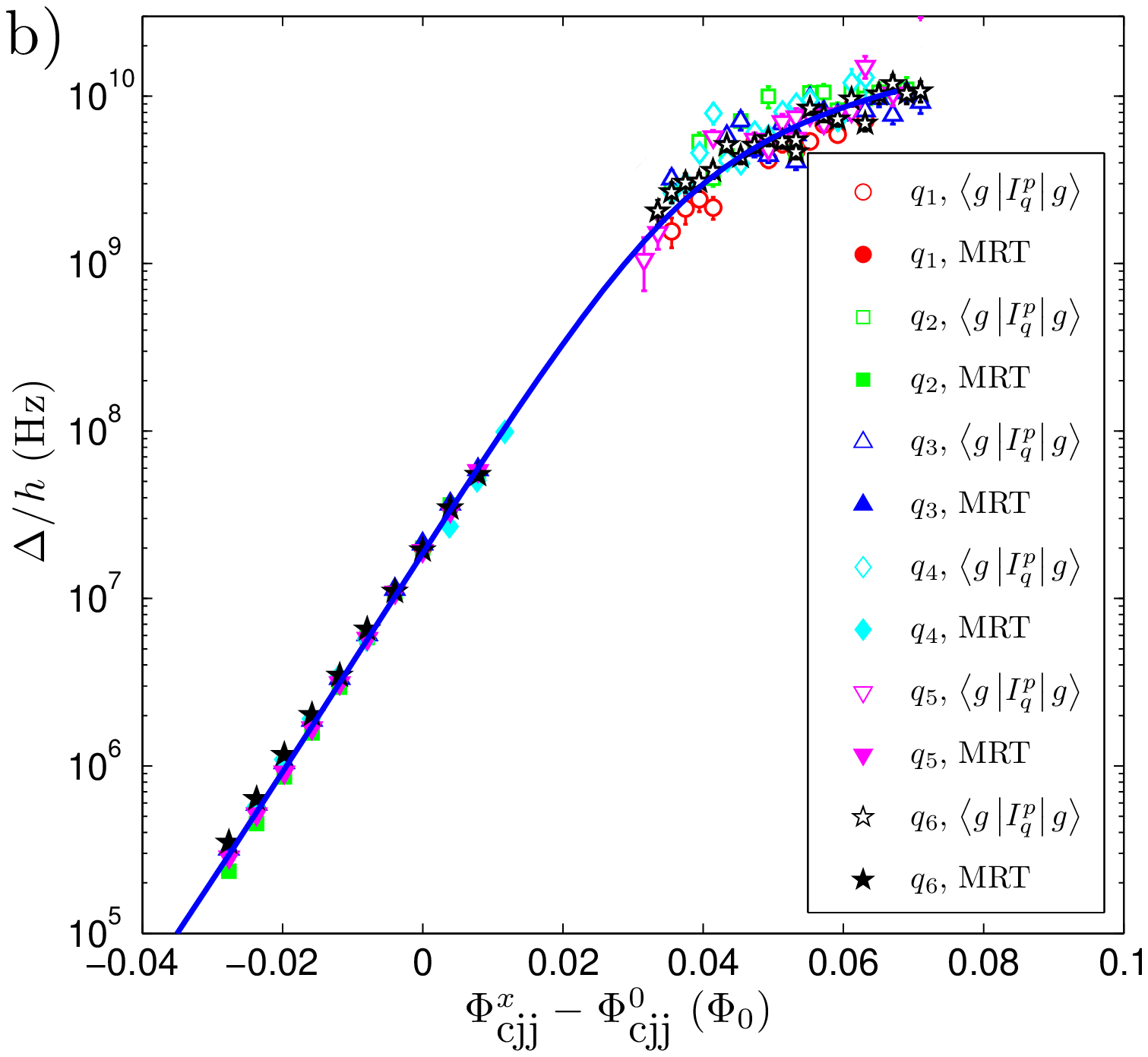}
\caption{(Color online) a) $\left|I_q^p\right|$ and b) $\Delta/h$ as a function of synchronized CJJ bias.  $\Delta/h$ from the 2-qubit and MRT measurement procedure are denoted as $\bra{g}I_q^p\ket{g}$ and MRT in the legend, respectively.  Solid curves are theoretical predictions using the mean device parameters quoted in Table \ref{tab:fit}.}
\label{fig:IpAndDelta}
\end{figure}

Measurements of the CJJ bias dependence of $\left|I_q^p\right|$ and $\Delta$ are shown in Fig.~\ref{fig:IpAndDelta}.  Here, we have shifted the CJJ bias for each qubit by a unique $\Phi_{\text{cjj}}^0$ (as summarized in Table \ref{tab:fit}) chosen such that $\Delta(\Phi_{\text{cjj}}^0)/h=10^7\,$Hz.  It can be seen that the six sets of $\left|I_q^p\right|$ data lie atop one another to within the measurement uncertainty over the range of $\Phi_{\text{cjj}}^x-\Phi_{\text{cjj}}^0$ for which $\Delta$ varies by five orders of magnitude.  The $\Delta$ data show reasonable synchronization, albeit the results for $q_2$ show higher $\Delta$ in the coherent regime and slightly faster exponential decay as a function of $\Phi_{\text{cjj}}^x$ at small $\Delta$.  Otherwise, the values of $\Delta$ from the other 5 qubits are synchronized to within $20\%$ over the range of $\Phi_{\text{cjj}}^x-\Phi_{\text{cjj}}^0$ shown.  The 2-qubit method for extracting $\Delta$ proved particularly susceptible to corruption by low frequency flux noise.  For these qubits, drift measurements of the type reported in Ref.~\cite{Drift} revealed $1/f$ noise spectral densities with a mean amplitude $\sqrt{S_{\Phi}(1\,\text{Hz})}=14\pm2\,\mu\Phi_0/\sqrt{\text{Hz}}$.  Efforts to refine the 2-qubit method and to reduce $1/f$ noise in our devices are ongoing. 

The results of simultaneously fitting $\left|I_q^p\right|(\Phi_{\text{cjj}}^x)$ and $\Delta(\Phi_{\text{cjj}}^x)$ using the eigenstates of Eq.~\ref{eqn:2JHeff} for each qubit are summarized in Table \ref{tab:fit}.  The quality of the fits proved most sensitive to $I_c$ and comparably less sensitive to the choice of $L_q$ and $C_p$.  It is probable that fabrication variations between Josephson junctions, roughly $\pm1\%$ of the target $I_c$, are the prime source of inter-qubit variability on this particular chip.  The theoretical predictions for $\left|I_q^p\right|(\Phi_{\text{cjj}}^x)$ and $\Delta(\Phi_{\text{cjj}}^x)$ using the mean device parameters listed at the bottom of Table \ref{tab:fit} are shown in Fig.~\ref{fig:IpAndDelta}.

\begin{table}
\begin{tabular}{|c|c|c|c|c|} \hline
Qubit & $\Phi_{\text{cjj}}^0$ (m$\Phi_0$) & $L_q$ (pH) & $C_p$ (fF) & $I_c$ ($\mu$A) \\ \hline\hline
1 & $-789\pm5$ & $200\pm2$ & $56\pm1$ & $2.58\pm0.01$ \\
2 & $-774\pm5$ & $202\pm2$ & $56\pm1$ & $2.65\pm0.01$ \\
3 & $-781\pm5$ & $200\pm2$ & $57\pm1$ & $2.63\pm0.01$ \\
4 & $-784\pm5$ & $202\pm2$ & $55\pm1$ & $2.59\pm0.01$ \\
5 & $-777\pm5$ & $200\pm2$ & $56\pm1$ & $2.65\pm0.01$ \\
6 & $-785\pm5$ & $202\pm2$ & $54\pm1$ & $2.59\pm0.01$ \\ \hline
Mean & $-782\pm12$ & $201\pm1$ & $56\pm1$ & $2.62\pm0.03$ \\ \hline
\end{tabular}
\caption{\label{tab:fit} Relative CJJ bias shifts $\Phi_{\text{cjj}}^0$ and device parameters obtained by simultaneously fitting $\left|I_q^p\right|(\Phi_{\text{cjj}}^x)$ and $\Delta(\Phi_{\text{cjj}}^x)$.}
\end{table}

{\it Conclusions:} A method for synchronizing the properties of multiple coupled CJJ rf-SQUID flux qubits with a small spread of device parameters due to fabrication variations has been demonstrated.  Both theory and experiment indicate that the application of a custom-tuned flux bias to each qubit CJJ loop is sufficient to compensate for $\pm1\%$ differences in critical current.  This strategy may prove to be an important step in the development of practical adiabatic quantum information processors.

We thank J.~Hilton, G.~Rose, P.~Spear, A.~Tcaciuc, F.~Cioata, E.~Chapple, C.~Rich, C.~Enderud, B.~Wilson, M.~Thom, S.~Uchaikin and M.H.S.~Amin. Samples were fabricated by the Microelectronics Laboratory of the Jet Propulsion Laboratory,
operated by the California Institute of Technology under a contract
with NASA.  S.Han was supported in part by NSF Grant No. DMR-0325551.



\begin{thebibliography}{99}


\bibitem{SCQubits} John Clarke and Frank K. Wilhelm, Nature {\bf 453}, 1031 (2008), and references therein.

\bibitem{Noise}  R.W. Simmonds \emph{et al.}, Phys. Rev. Lett.
\textbf{93}, 077003 (2004); J.M. Martinis \emph{et al.}, Phys. Rev. Lett. \textbf{95}, 210503 (2005); R.C. Bialczak \emph{et al.}, Phys. Rev. Lett. \textbf{99}, 187006 (2007); O.~Astafiev, Y.A. Pashkin,Y.~Nakamura, T.~Yamamoto and J.S. Tsai, Phys. Rev. Lett. \textbf{93}, 267007 (2004); J. A.~Schreier \emph{et al.}, Phys. Rev. B {\bf 77}, 180502(R) (2008).

\bibitem{Drift} T.~Lanting \emph{et al.}, Phys. Rev. B {\bf 79}, 060509(R) (2009).

\bibitem{Sensitivity} R.C.~Ramos \emph{et al.}, IEEE Trans. Appl. Supercond. {\bf 11}, 998 (2001); Jens Koch \emph{et al.}, Phys. Rev. A {\bf 76}, 042319 (2007); J.E.~Mooij \emph{et al.}, Science {\bf 13}, 285 (1999); F.G. Paauw, A. Fedorov, C.J.P.M. Harmans and J.E. Mooij, Phys. Rev. Lett. {\bf 102}, 090501 (2009).

\bibitem{CJJ}  {S.~Han, J.~Lapointe and J.E.~Lukens, Phys. Rev. Lett. {\bf 63}, 1712 (1989); S.~Han, J.~Lapointe and J.E.~Lukens, Phys. Rev. Lett. {\bf 66}, 810 (1991).}

\bibitem{Farhi} E.~Farhi \emph{et al.}, Science {\bf 292}, 472 (2001);

\bibitem{AQC} D.~Aharonov, W.~van Dam, J.~Kempe, Z.~Landau, and S.~Lloyd,
SIAM Journal of Computing {\bf 37},166 (2007); Jacob D. Biamonte and Peter J. Love, Phys. Rev. A {\bf 78}, 012352 (2008); Jingfu Zhang \emph{et al.}, Phys. Rev. A {\bf 79}, 012305 (2009).

\bibitem{architecture} W.M. Kaminsky and S.~Lloyd, in \textit{Quantum Computing and Quantum Bits in Mesoscopic Systems}, MQC$^2$ (Kluwer Academic, New York USA, 2003).


\bibitem{Boros} E. Boros, P.L. Hammer and G. Tavares,  J. Heuristics {\bf 13}, 99 
(2007).


\bibitem{Landaubook} L.~D. Landau and E.~M. Lifshitz, {\it Quantum Mechanics}, Vol. 3, 3rd ed., 179, (Butterworth-Heinemann, 1977).

\bibitem{CJJCoupler}  A. Maassen van den Brink, A.J. Berkley, and M.~Yalowsky, New J. Phys. \textbf{7}, 230 (2005).

\bibitem{MRT} M.H.S. Amin and D.V. Averin, Phys. Rev. Lett. {\bf 100}, 197001 (2008); R.~Harris \emph{et al.}, Phys. Rev. Lett. 101, 117003 (2008),

\bibitem{IPHTDelta}  A.~Izmalkov \emph{et al.}, Phys. Rev. Lett. 101, 017003 (2008).

\end{thebibliography}
\end{document}